\documentclass[nofootinbib,twocolumn,showpacs,preprintnumbers,amsmath,amssymb]{revtex4}

\usepackage[dvips]{graphicx}
\usepackage{slashed}

\usepackage{footmisc}

\newcommand{\rig}{\rightarrow}

\newcommand{\be}{\begin{eqnarray*}}
\newcommand{\ee}{\end{eqnarray*}}
\newcommand{\gl}[1]{(\ref{#1})}

\newcommand{\bee}{\begin{eqnarray}}
\newcommand{\eee}{\end{eqnarray}}
\newcommand{\beeq}{\begin{equation}}
\newcommand{\eeeq}{\end{equation}}
\newcommand{\gev}{~{\rm{GeV}}}

\def\lesssim{\mathrel{\raisebox{-.6ex}{$\stackrel{\textstyle<}{\sim}$}}}

\begin{document}

\preprint{FTUV-10-0612\;\;KA--TP--16--2010\;\;SFB/CPP--10--50}

\title{QCD corrections to non-standard WZ+jet production\\with leptonic decays at the LHC} 

\author{Francisco~Campanario}
\email{francam@particle.uni-karlsruhe.de}
\affiliation{Institute for Theoretical Physics, KIT, 76128 Karlsruhe, Germany}
\author{Christoph~Englert}
\email{c.englert@thphys.uni-heidelberg.de}
\affiliation{Institute for Theoretical Physics, KIT, 76128 Karlsruhe, Germany}
\affiliation{Institute for Theoretical Physics, Heidelberg University, 69120 Heidelberg, Germany}
\author{Michael~Spannowsky}
\email{mspannow@uoregon.edu}
\affiliation{Institute for Theoretical Science, University of Oregon, Eugene Oregon 97403, USA}

\begin{abstract}
We discuss the impact of anomalous WW$\gamma$ and WWZ couplings on WZ+jet production at next-to-leading order QCD, 
including full leptonic decays of the electroweak gauge bosons. While the inclusive hadronic cross sections do not exhibit any particular
sensitivity to anomalous couplings once the residual QCD scale uncertainties are taken into account, the 
transverse momentum distributions show substantial deviations from the Standard Model, provided that
the anomalous vertices are probed at large enough momentum transfers.
\end{abstract} 


\maketitle        
%
%
%
\section{Introduction}
Searches for anomalous trilinear electroweak couplings at 
hadron colliders is an active field of research to discover
new interactions beyond the well-established Standard Model of particle physics (SM), see, e.g.,~\cite{aba:2010qn} for
very recent D$\slashed{\rm{0}}$ results.  Crucial for revealing
modifications of the electroweak sector is, without any doubt, adequate experimental and theoretical precision of measurement and simulation, respectively.
Since corrections from Quantum
chromodynamics (QCD) are typically sizable at hadron colliders, they have to be properly included in phenomenological studies 
in order not to misinterpret QCD dynamics for beyond-the-SM (BSM) interactions when data is analyzed.
Hence, an important step towards resolving these QCD-related issues is studying
important production processes at least to next-to-leading order (NLO) QCD precision. 

The NLO QCD corrections to anomalous $WZ$ and $W\gamma$ production have already been discussed in the literature in detail, 
e.g. in Refs.~\cite{Baur:1993ir,Dobbs:2005ev}. At the LHC, with its
large center-of-mass energy of expected 14 TeV, additional jet radiation from accessing gluon-induced channels with
large parton luminosities is substantially important for these processes, which are exclusively quark-induced at LO. 
Moreover, the additional parton emission
represents a theoretically indispensable real emission contribution to the infrared-safe NLO QCD diboson cross section 
via the Kinoshita-Lee-Nauenberg theorem~\cite{Kinoshita:1962ur}. By probing the gluon-induced channels 
at small momentum fractions with respect to the incoming protons, the one-jet-contribution becomes comparable to the leading order 
(zero jet) cross section for typical selection criteria\footnote{We
refer to the considered processes 
$pp\rig \ell^-\bar\nu_\ell \ell'^+\ell'^-+~\rm{jet}$ and $pp\rig 
\ell^+\nu_\ell \ell'^+\ell'^-+~\rm{jet}$, where $\ell,\ell'$ denote light but distinct lepton flavors, as
$W^\pm Z+{\rm{jet}}$ production in this paper.},
\bee
\label{crosssecs}
{\sigma(\hbox{$pp\rig 3~\rm{leptons}+\slashed{p}_T+{\rm{jet}}$}) \over
\sigma(\hbox{$pp\rig 3~\rm{leptons}+\slashed{p}_T$})}\simeq 1\,.
\eee
Hence, for inclusive measurements, the perturbative uncertainty of NLO $WZ$ production is heavily affected by the one jet contribution, which is leading order (LO) in the 
strong coupling-expansion.

Since the gluon-induced real emission contributions kinematically obstruct anomalous
coupling measurements, one often reverts to jet vetos, at least in some way, when performing phenomenological studies, e.g. in Ref.~\cite{Dobbs:2005ev}. 
Hence, considering exclusive NLO production~\cite{Baur:1993ir},
one is able to increase the sensitivity of the $WZ$ cross section to anomalous
couplings, which we define to be $\sigma(\hbox{non-SM})/\sigma(\hbox{SM})$ in the following.
From a perturbative QCD point of view,
however, this strategy does not necessarily give rise to improved QCD precision of total cross sections, even if improved renormalization and 
factorization scale dependencies might superficially suggest so. To arrive at seemingly stable NLO results, one subtracts a dominant and unreliable 
LO-$\alpha_s$ contribution. Since this very contribution is sensitive to the applied hadronic cuts, one is able, given a relation analogous to Eq.~\gl{crosssecs}, 
to tune the real emission contribution to milden the factorization and renormalization
scale dependencies at the integrated cross section level~\footnote{Indeed, this is the case for e.g. $W\gamma$+jet production, where a
more thorough investigation reveals sophisticated cancellations of scale dependencies for the NLO exclusive cross sections, rendering the perturbative stability unreliable.}.
This signalizes the necessity to also take into account the differential QCD corrections to diboson production in association
with a hard jet, to add NLO precision to the veto performance. The quantitative dependence of the QCD corrections on the anomalous couplings in this context
supplements the necessary additional information to recent theoretical leading order analysis \cite{Eboli:2010qd}.

Eq.~\gl{crosssecs} also raises the question whether there is enough sensitivity to anomalous couplings left at larger inclusive rates, which could supplement
the powerful ``traditional'' measurement strategies, that exploit the well-known classical radiation zeros in the $q\bar Q \rig {W}\gamma, {WZ}$
amplitudes in the SM~\cite{Baur:1994ia}.
Given Eq.~\gl{crosssecs}, 
the sensitivity to anomalous couplings in inclusive phenomenological analysis crucially depends on the one jet-inclusive cross section. 
A reliable assessment of the anomalous couplings' impact is then only possible if QCD corrections are properly included as they have turned out to be
particularly sizable for the diboson+jet cross sections in Refs.~\cite{Dittmaier:2007th,Campanario:2009um, Binoth:2009wk,wztoapp}. 

To contribute to resolving the above issues, we report in this paper on the calculation of non-standard $WZ+{\rm{jet}}$ production at NLO QCD.
We compute the $WZ+{\rm{jet}}$ cross sections and distributions as functions of the anomalous couplings' parameters 
by means of a fully-flexible Monte Carlo program that has been designed for the purpose of this work.
Full leptonic decays of the $W$ and $Z$ bosons are included throughout. The code will be publicly available with an upcoming update of {\sc Vbfnlo}~\cite{Arnold:2008rz},
and is also capable of performing Tevatron calculations.
We give details on the computation in Sec.~\ref{sec:details}, and Sec.~\ref{sec:res} is devoted to the numerical results. In particular, we compare the impact of anomalous
couplings to the scale uncertainty of the SM cross section for inclusive selection cuts and for a cut choice that mimics genuine $WZ$ events. For the latter, we discuss the
anomalous couplings' impact on the differential cross sections in more detail. 
Sec.~\ref{sec:conc} closes with a summary of this work.

\section{Details of the calculation}
\label{sec:details}
The strategy and the methods we apply to perform the NLO calculation 
in a fast and numerically stable way have already been presented in detail in Refs.~\cite{Campanario:2009um,wztoapp}. We
therefore only sketch the numerical implementation to make this work self-consistent. 
To arrive at the numerical results of Sec.~\ref{sec:res}, we use a modified version of the Monte Carlo code of Ref.~\cite{wztoapp}, which builds upon
the \textsc{Vbfnlo} framework~\cite{Arnold:2008rz}. We evaluate the LO matrix elements for $pp \rig 3~\hbox{leptons}+\slashed{p}_T+~\hbox{jet}$ 
at ${\mathcal{O}}(\alpha_s\alpha^4)$
via \textsc{Helas} routines~\cite{Murayama:1992gi}, which are set up with \textsc{MadGraph}~\cite{Alwall:2007st}.
The virtual and real emission matrix elements are computed numerically using in-house routines 
along the lines described in Refs.~\cite{Campanario:2009um,wztoapp}. Algebraic bookkeeping of the infrared singularities is thereby performed by 
applying the method of Catani and Seymour~\cite{Catani:1996vz}. Throughout the computation, the anomalous trilinear couplings are 
included via purpose-built \textsc{Helas} routines, which we design using an in-house framework that employs {\sc{FeynRules}}~\cite{Christensen:2008py}.
The anomalous trilinear vertices follow from the most general 
Lorentz-, ${\mathcal{CP}}$-, and QED-invariant effective operators up to dimension six~\cite{Baur:1993ir,Hagiwara:1986vm},
\beeq
\begin{split}
\label{anovertex}
\mathcal{L}_{{{WW}}\gamma} =& 
-ie \big[ W_{\mu\nu}^\dagger W^\mu A^\nu- W_\mu^\dagger A_\nu W^{\mu\nu} \\
&+\kappa_\gamma  W_\mu^\dagger W_\nu F^{\mu\nu} +{\lambda_\gamma\over m_W^2} W_{\lambda\mu}^\dagger W^\mu_\nu
F^{\nu\lambda}\big]\,
\end{split}
\eeeq
for the anomalous $WW\gamma$ vertex, and
\beeq
\begin{split}
\label{anovertexZ}
\mathcal{L}_{{{WWZ}}} = &
-ie \cot \theta_w \,\big[ g_1^Z\left( W_{\mu\nu}^\dagger W^\mu A^\nu- W_\mu^\dagger A_\nu W^{\mu\nu}\right)\\
&+\kappa_Z  W_\mu^\dagger W_\nu Z^{\mu\nu} +{\lambda_Z\over m_W^2} W_{\lambda\mu}^\dagger W^\mu_\nu
Z^{\nu\lambda}\big]\,
\end{split}
\eeeq
for the anomalous $WWZ$ vertex. In Eq.~\gl{anovertexZ} $\theta_w$ denotes the weak mixing angle. 
$W^\mu$, $A^\mu$, and $Z^\mu$ denote the $W$, the photon and the $Z$ boson. The field strength
tensors are $F^{\mu\nu} = \partial^\mu A^\nu - \partial^\nu A^\mu$, $Z^{\mu\nu} = \partial^\mu Z^\nu - \partial^\nu Z^\mu$, and
$W^{\mu\nu} = \partial^\mu W^\nu - \partial^\nu W^\mu$, as usual.
We do not include ${\cal{CP}}$-violating operators since they are already heavily constrained by measurements of the neutron's electric dipole moment (Ref.~\cite{Baur:1993ir,Amsler:2008zzb}).
We thus can focus on the $W^-Z$+jet production for brevity, and our findings generalize to $W^+Z$+jet production in a rather straightforward fashion. The $W^+Z$+jet cross sections
are, however, larger by a factor 1.5 due to testing the protons' valence quarks. 

We apply the gauge constraints, which were used for the combined analysis of LEP data in Ref.~\cite{Alcaraz:2006mx},
\bee
\kappa_Z=g^Z_1 - (\kappa_\gamma -1)\tan^2\theta_w\,, \quad \lambda_Z=\lambda_\gamma\,.
\eee
This analysis also gives rise to the currently most stringent bounds on anomalous trilinear couplings,
\begin{figure}[b]
\begin{center}
\includegraphics[width=0.46\textwidth]{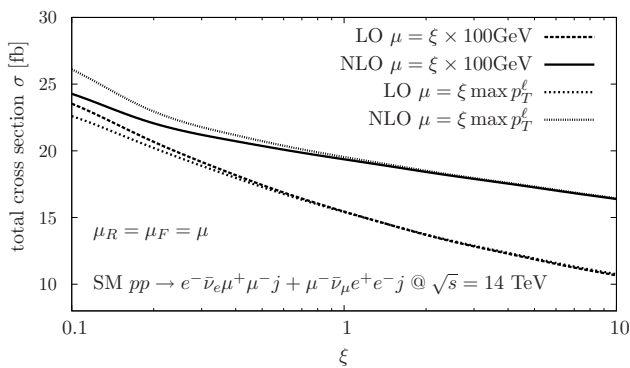}
\end{center}
\vspace{-0.4cm}
\caption{\label{fig:scaldep}
Scale dependence of the SM cross section at LO and NLO for the set up described in Sec.~\ref{sec:res}. In addition
to the fixed-scale variation, we also plot the variation for renormalization and factorization scales chosen to be
the maximum transverse momentum of the charged leptons. 
Varying $\mu$ by a factor two around $\xi=1$ reduces the uncertainties
from $24\%$ ($23\%$) at LO to $9\%$ ($10\%$) at NLO for $\mu=\xi\times 100\gev$ ($\mu=\xi\max p_T^\ell$).}
\end{figure}
\begin{multline}
\label{eq:bounds}
g_1^Z=0.991^{+0.022}_{-0.021}\,,\;
\kappa_\gamma = 0.984^{+0.042}_{-0.047}\,,\; \\
\lambda_\gamma = - 0.016^{+0.021}_{-0.023}\,,
\end{multline}
\noindent at 68\% confidence level. Most constraining bounds at hadron colliders have only recently
been obtained from direct measurements of the $WWZ$ vertex at the Fermilab D$\slashed{\rm{O}}$~\cite{aba:2010qn}.

Unitarity requires the anomalous parameters $\{g_1^Z,\kappa_Z,\lambda_Z,\kappa_\gamma,\lambda_\gamma\}$ to be understood
as low-energy form factors. Their precise momentum dependence is
determined by the BSM Lagrangian. 
Avoiding any particular assumption about how the BSM interactions might look like (except for participation in
SM interactions of some kind), we use the conventional dipole parametrization of the form factors, Ref.~\cite{Baur:1993ir}. Introducing new
parameters, which rephrase the modifications of Eqs.~\gl{anovertex} and \gl{anovertexZ} around the SM Lagrangian,
\bee
(\Delta g_1^Z,\Delta \kappa_Z, \Delta\kappa_\gamma) = ( g_1^Z,\kappa_Z,\kappa_\gamma) - 1\,,
\eee
the dipole profile is given by
\beeq
\label{eq:param}
(\Delta g_1^Z,\Delta\kappa_Z,\lambda_Z,\Delta\kappa_\gamma,\lambda_\gamma) = { (\Delta g_1^{Z, 0},\Delta\kappa_Z^0,\lambda_Z^0,
\Delta\kappa_\gamma^0,\lambda^0_\gamma) 
\over \big(1+m_{WZ}^2/\Lambda^2\big)^{2}}\,,
\eeeq
where $m_{WZ}$ denotes the invariant mass of the decaying $WZ$ or $W\gamma$ pair, and $\Lambda$ is the scale where new physics enters the picture. 
The momentum dependence of Eq.~\gl{eq:param} guarantees vanishing contributions from phase space regions where the naive momentum-independent parametrization 
violates unitarity. In addition, given
the steeply falling parton luminosities for large parton momentum fractions, a well-behaved theory yields vanishing contributions in this particular region, so that
the parametrization of Eq.~\gl{eq:param} is adequate for phenomenological studies.
In the following, we choose $\Lambda=2~\rm{TeV}$, which is also a benchmark point of various experimental studies (e.g. in Ref.~\cite{Dobbs:2005ev,aba:2010qn}).
\section{Numerical results}
\label{sec:res}
For the numerical results, we use CTEQ6M parton distributions~\cite{Pumplin:2002vw} and the CTEQ6L1 set at LO. 
We choose $m_{Z}=91.188~\rm{GeV}$, $m_{W}=80.419~\rm{GeV}$ and $G_F=1.16639\times 10^{-5}~\textnormal{GeV}^{-2}$ as electroweak input parameters and derive the electromagnetic 
coupling $\alpha$ and the weak mixing angle from Standard Model-tree level relations. The LO and NLO running of the strong coupling $\alpha_s$ 
is determined by $\alpha_s^{\rm{LO}}(m_{Z})=0.130$ and $\alpha_s^{\rm{NLO}}(m_{Z})=0.118$ with five active flavors, respectively.
We fix the center-of-mass energy to $14~\rm{TeV}$ for LHC collisions,
and we sum over the light lepton flavors in the final state, which we
treat as massless.

Additionally, we neglect processes subject
to Pauli-interference, i.e. we quote results for $pp \rig e^-\bar \nu_e\mu^+\mu^- j + X$ {\emph{plus}} $pp \rig \mu^-\bar \nu_\mu e^+e^- j + X$  in case 
of $W^-Z+{\rm{jet}}$ production \footnote{Since Pauli-interference
is a small effect for our process, we could as well multiply our results for the total cross section and
differential distributions by a global factor 1.5 to account
for the processes with identical final state lepton flavors.}. Lowering the available LHC energy 
to 7 TeV yields a too low rate to be of phenomenological importance given the scheduled integrated luminosity of 1 fb$^{-1}$.
Since anomalous couplings alter the high-energy phenomenology,  $WZ+{\rm{jet}}$ production from Tevatron collisions does not yield any notable sensitivity in the allowed parameter range.
\begin{figure}[t]
\centering
\includegraphics[scale=0.52]{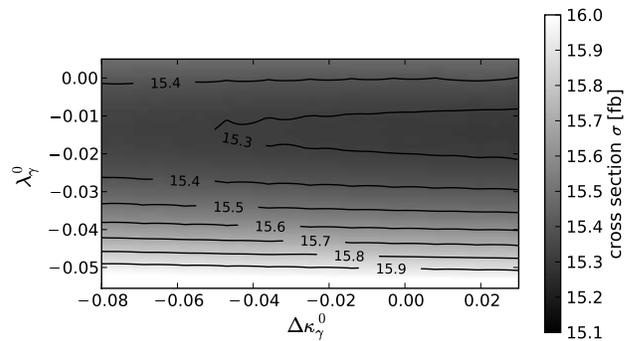}
\caption{\label{fig:lototalxsecs}
Total LO cross section contours in fb for $W^-Z$+jet production for the cut and parameter choices as described in the text. We use a fixed-scale
choice $\mu_R=\mu_F=100\gev$. Additionally, we choose
$\Delta g^{Z,0}_1=0$.}
\vspace{-0.4cm}
\end{figure}
\begin{figure}[t]
\centering
\includegraphics[scale=0.52]{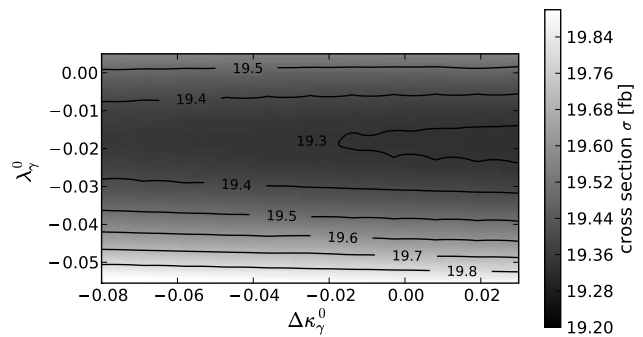}
\caption{\label{fig:nlototalxsecs}
Total NLO cross section contours in fb for $W^-Z+{\rm{jet}}$ production analogous to Fig.~\ref{fig:lototalxsecs}.}
\vspace{-0.4cm}
\end{figure}
\begin{figure}[t!]
\centering
\includegraphics[scale=0.52]{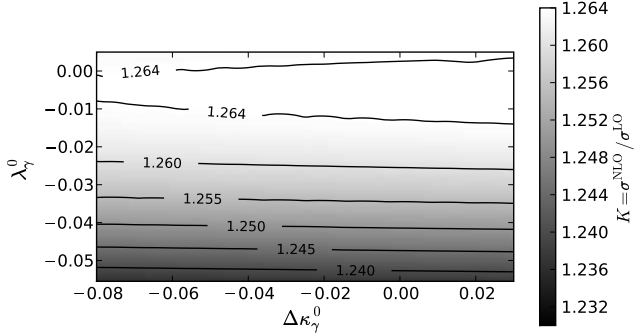}
\caption{\label{fig:totalk}
Total $K$ factor contours for $W^-Z+{\rm{jet}}$ production analogous to Fig.~\ref{fig:lototalxsecs}.}
\vspace{-0.4cm}
\end{figure}

The CKM matrix is assumed to be diagonal, which is an excellent approximation for LHC cross sections~\cite{wztoapp}.
Jets are recombined via the algorithm of~\cite{Catani:1993hr} from partons which fall in the pseudorapidity range of $|\eta|\leq 5$. 
The jet resolution parameter is chosen to be $D=0.7$, and 
the jets are required to lie in the rapidity range accessible to the detector 
\bee 
|y_j | \leq 4.5\,.
\eee
 For the charged leptons, we request 
\bee
|\eta_\ell|\leq 2.5\,.
\eee
We choose a very inclusive cut set-up in order to
analyze the anomalous couplings' impact on the phenomenology of  $WZ+{\rm{jet}}$ production over
a wide range of accessible phase space at the LHC. We require
\beeq
p_T^j\geq 30\gev\,,\quad p_T^\ell\geq 25\gev\,, \quad \slashed{p}_T \geq 25\gev\,,
\eeeq
and, for the jet-lepton and lepton-lepton separation in the azimuthal angle--pseudorapidity plane
\beeq
R_{  \ell'\ell} = \sqrt{(\Delta\phi_{\ell'\ell}^2 + \Delta\eta_{\ell'\ell}^2)}\geq 0.2\,,\quad R_{  j\ell}  \geq 0.2\,, 
\eeeq
for all charged leptons and reconstructed jets.
\begin{figure}[t!]
\includegraphics[scale=0.52]{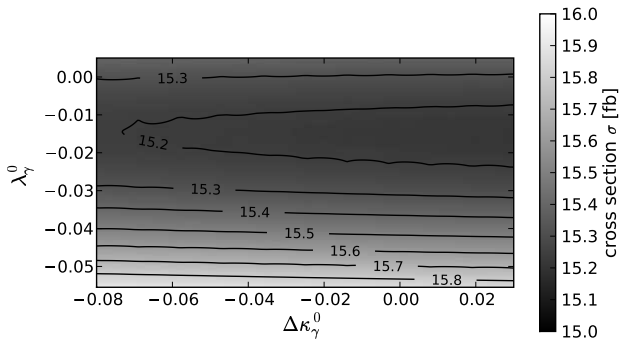}
\caption{\label{fig:diffptscan}
Total NLO cross section contours in fb for $W^-Z+{\rm{jet}}$ production for the cut and parameter choices as described in the text, with the additional requirement of
Eq.~\gl{eq:ellcriter}. We again choose $\Delta g^{Z,0}_1=0$. We choose $\mu_R=\mu_F=\max p_T^\ell$.}
\vspace{-0.4cm}
\end{figure}
\begin{figure}[t!]
\centering
\includegraphics[scale=0.52]{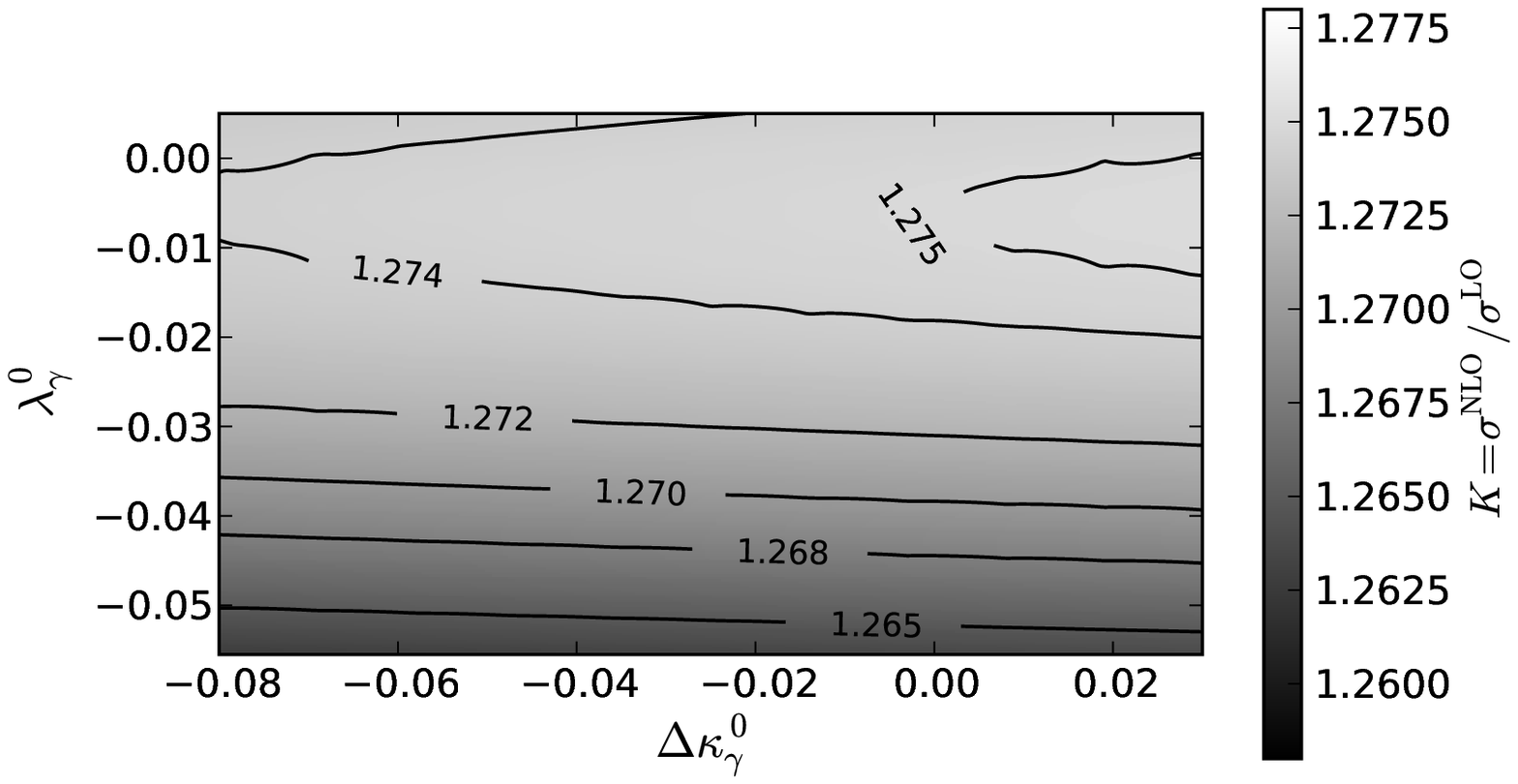}
\caption{\label{fig:totalkptscan}
Total $K$ factor contours for $W^-Z+{\rm{jet}}$ production analogous to Fig.~\ref{fig:diffptscan}.}
\vspace{-0.4cm}
\end{figure}
\begin{figure}[b!]
\begin{center}
\includegraphics[width=0.46\textwidth]{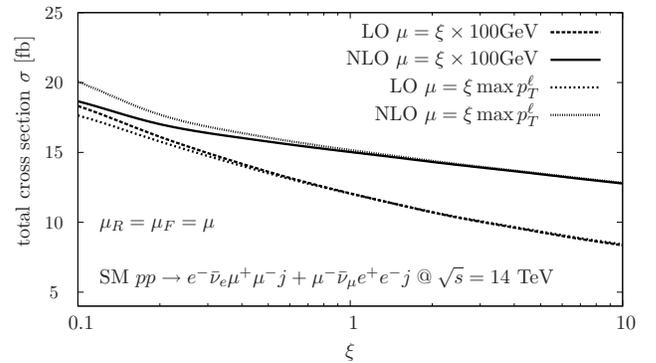}
\end{center}
\vspace{-0.4cm}
\caption{\label{fig:scaldep2}
Scale dependence of the SM cross section at LO and NLO for the inclusive set-up described in Sec.~\ref{sec:res} with
the additional requirement of Eq.~\gl{eq:ellcriter}.
Analogous to Fig.~\ref{fig:scaldep}, the scale uncertainty is reduced from 
$24\%$ ($23\%$) at LO to $10\%$ ($11\%$) at NLO for $\mu=\xi\times 100\gev$ ($\mu=\xi \max p_T^\ell$).}
\vspace{-0.4cm}
\end{figure}

\begin{figure}[t]
\begin{center}
\includegraphics[width=0.46\textwidth]{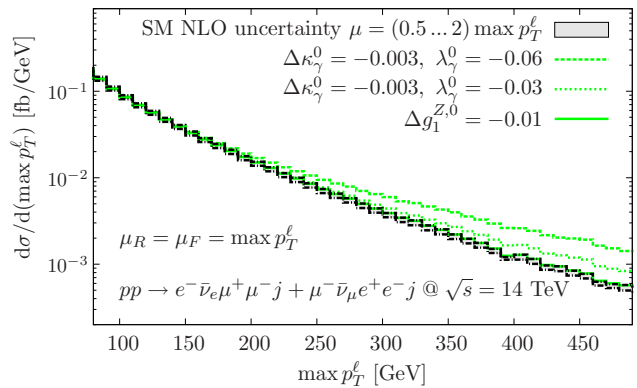}
\end{center}
\vspace{-0.4cm}
\caption{\label{fig:diffpt}
NLO inclusive distribution of the leptons' maximum transverse momentum $\max p_T^\ell$ for various values of the anomalous couplings. 
The anomalous parameters which are not quoted are chosen to be zero. The solid line lies within the SM uncertainty band, and the plotted distributions 
are affected by scale uncertainty bands of equal width. For the shown distributions we have applied the cut of Eq.~\gl{eq:ellcriter}.}
\vspace{-0.4cm}
\end{figure}

In Figs.~\ref{fig:lototalxsecs} and~\ref{fig:nlototalxsecs}, we representatively scan the NLO 
cross section for the fixed-scale choice $\mu_R=\mu_F=100\gev$ 
over the allowed parameter range for the anomalous couplings of Eq.~\ref{eq:bounds} with the additional requirement $\Delta g_1^{Z,0}=0$. The total $K$ factor contours,
\bee	
K={\sigma^{\rm{NLO}} \over \sigma^{\rm{LO}}}\,,
\eee
are depicted in Fig.~\ref{fig:totalk}. In particular, the anomalous production cross section mostly depends on the dimension six operators of 
Eqs.~\gl{anovertex} and \gl{anovertexZ}, which are not present in the SM Lagrangian. 
The QCD corrections are sizable over the whole range of anomalous parameters allowed by LEP measurements. 
They are especially important for values close to the SM. This follows from the QCD corrections, which turn out to be particularly large around the $p_T$ 
thresholds and decreasingly important for larger transverse momentum values in the $p_T$ distributions' tails. The corrections for our cut choices 
are relatively large for small values of $p^\ell_T\lesssim 150\gev$ ($K \simeq 1.3$), where the distributions are unaffected 
by the anomalous couplings (Fig.~\ref{fig:diffpt}). The decreasing $K$ factor when approaching
the boundaries in Fig.~\ref{fig:totalk} can be understood accordingly: 
For large values of the anomalous couplings, the cross section is altered already at LO with respect to the SM, so that its increase from
QCD corrections is less significant. This implies that the NLO QCD corrections decrease the sensitivity to anomalous couplings.

Confronting the non-SM cross sections with the perturbative uncertainty of ${\cal{O}}(10\%)$ for SM-like production, Fig.~\ref{fig:scaldep}, it is apparent that
for our inclusive cuts the impact of the anomalous couplings entirely drowns in the residual QCD scale uncertainty at NLO. In Fig.~\ref{fig:scaldep}, we also
plot the scale variation for the intrinsic scale $\max p_T^\ell$, which is 
via Eq.~\gl{eq:param}, related to the characteristic scale of the anomalous couplings.
Even for hard events with $\max p_T^\ell\gtrsim 150\gev$, the impact of the anomalous
parameters is not apparent, which also explains the small percent-level deviations of the NLO cross section over the allowed parameter range in
Fig.~\ref{fig:nlototalxsecs}. The vanishing sensitivity arises from only small
momentum transfers in the anomalous trilinear vertices due to our selection criteria: the jets recoil predominantly against
the collinear $WZ$ pair, which is an anomalous couplings-insensitive kinematical configuration. A straightforward way to induce considerably larger momentum transfers while reducing the
contributions from anomalous couplings-insensitive graphs, where both the $W$ boson and the $Z$ boson couple to the quark legs, is therefore requiring a large separation
of the identically charged decay leptons. Note, that this reflects the kinematics of exclusive diboson production with the $W$ and $Z$ recoiling against each
other at LO. A convenient choice is
\bee
\label{eq:ellcriter}
R_{\ell^\pm \ell'^\pm} \geq 1.5\,.
\eee
While the cross section decreases by approximately 20\%, Fig.~\ref{fig:diffptscan} reveals cross section deviations due to anomalous couplings of order 5\% compared
to the SM. Although this is still comparable to the cross section's scale dependence plotted in Fig.~\ref{fig:scaldep2}, its increase compared to the SM entirely results
from the large-$p_T^\ell$ phase space region, Fig.~\ref{fig:diffpt}. In this region, we find substantial deviations in the $\max p_T^\ell$ shape,
which can be well outside the SM scale uncertainty for larger values of the anomalous couplings. 
Hence, provided a sufficiently large momentum transfer in the anomalous vertices, the $WZ+{\rm{jet}}$ production 
which is vetoed in $pp\rig {WZ}$ analysis, exhibits potential sensitivity to anomalous couplings via fits to the $p_T$ distributions. However, the total cross section 
becomes tiny and the sensitivity to anomalous couplings decreases by including the NLO corrections as can be inferred from Figs.~\ref{fig:totalk} and~\ref{fig:totalkptscan}.

To this end, it is worthwhile to briefly comment on one-jet exclusive $WZ+{\rm{jet}}$ production. In particular, it has been shown
that the exclusive diboson+jet cross sections are seemingly stable.
In ~\cite{wztoapp}, however, it was also shown that applying an additional fixed-$p_T$ jet veto on the second reconstructed jet at NLO yields a poor perturbative reliability.
In particular, applying such a veto can lead to negative bins already for modest scales of about 100 GeV. 
Indeed, this is the phase space region where anomalous couplings give rise to well-pronounced 
deviations from the SM phenomenology (Eq.~\gl{eq:param}) at LO. Hence, perturbation theory forces us to consider inclusive production to give reliable predictions at NLO; 
for our inclusive cut choices with respect to hadronic activity our numerical results are not affected by the mentioned pathologies and can be considered
 stable except for the residual scale uncertainty of about 10\%.
 
\section{Summary}
\label{sec:conc}
We have discussed the impact of anomalous trilinear couplings on NLO QCD $W^-Z+{\rm{jet}}$ production at the LHC, including leptonic decays. 
We do not find any significant deviations of differential cross sections, unless we induce sufficiently large momentum transfers in the trilinear vertices. This can 
be realized by requiring back-to-back $WZ$ pairs.
The resulting modifications are characterized by large transverse momenta, and are well-outside the 
SM scale uncertainty that is intrinsic to our NLO QCD computation for large values of the anomalous couplings in the allowed range by LEP.
In general we find that the differential cross sections' sensitivity to anomalous couplings decreases when including 
inclusive NLO corrections (Figs.~\ref{fig:totalk} and~\ref{fig:totalkptscan}), while the QCD corrections do not exhibit any particular
dependence on the anomalous parameters. Performing precise measurements in a full hadronic environment, for which our calculation
is relevant, requires a careful analysis, taking into account all systematic effects, ranging from showering to detector effects. 
This is clearly beyond the scope of our calculation.
Our Monte Carlo code will be publicly available with an upcoming update of {\sc Vbfnlo}. \\

{\bf{Acknowledgments}} ---
F.C. acknowledges partial support by FEDER and Spanish MICINN under grant
FPA2008-02878.
C.E. and M.S. thank the organizers of the {\sc{NobEl}} (Nordbadische Eliteuniversit\"aten) particle physics meetings Uli Nierste and Tilman Plehn.
C.E. also would like to thank the High Energy Physics group of the Institute for Theoretical Science at the University of Oregon for its hospitality during the time
when this work was completed and acknowledges partial support by ``KCETA Strukturiertes Promotionskolleg''. 
This research is partly funded by the Deutsche Forschungsgemeinschaft under 
SFB TR-9 ``Computergest\"utzte Theoretische Teilchenphysik'', and the Helmholtz alliance ``Physics at the Terascale''.

\end{document}